\documentclass[12pt,reqno]{amsart}

\usepackage{graphicx}
\graphicspath{{EPS files/}}
\usepackage{amssymb, amsmath, amsthm}
\usepackage[colorlinks=true,linkcolor=blue,citecolor=red]{hyperref}

\usepackage{tikz}
\usepackage{pgfplots} 
\pgfplotsset{width=10cm,compat=1.9}
\newtheorem{theorem}{Theorem}

\newtheorem{remark}{Remark}

\newcommand{\R}{\mathbb{R}}
\newcommand{\Z}{\mathbb{Z}}
\newcommand{\C}{\mathbb{C}}

\newcommand{\om}{\omega}

\newcommand{\mL}{\mathcal L}

\newcommand{\lm}{\lambda}
\newcommand{\gm}{\gamma}
\newcommand{\mc}{\mathcal}
\newcommand{\vp}{\varphi}

\begin{document}

\title[Stability of smooth solitary waves]{\bf Stability of smooth solitary waves \\ under the intensity--dependent dispersion}

\author{P.G. Kevrekidis}
\address[P.G. Kevrekidis]{Department of Mathematics and Statistics,
	University of Massachusetts, Amherst, MA 01003-4515, USA}

\author{D.E. Pelinovsky}
\address[D.E.Pelinovsky]{Department of Mathematics and Statistics, McMaster
	University, Hamilton, Ontario, Canada, L8S 4K1}

\author{R.M. Ross}
\address[R.M. Ross]{Department of Mathematics and Statistics,
	University of Massachusetts, Amherst, MA 01003-4515, USA}

\begin{abstract}
The cubic nonlinear Schr{\"o}dinger equation (NLS) in one dimension is
considered in the presence of an intensity-dependent dispersion term. We study bright solitary waves with  smooth profiles which extend from the limit where the dependence of the dispersion coefficient on the wave intensity is negligible to the limit where the solitary wave becomes singular due to vanishing dispersion coefficient. We analyze and numerically explore the stability 
for such smooth solitary waves, showing with the help of numerical approximations that the family of solitary waves becomes unstable 
in the intermediate region between the two limits, while being stable 
in both limits. This bistability, that has also been observed in other
NLS equations with the generalized nonlinearity, brings about interesting dynamical transitions from one stable branch to another stable branch, 
that are explored in direct numerical simulations of the NLS equation with the intensity-dependent dispersion term.
\end{abstract}

\date{\today}
\maketitle

\section{Introduction}

The cubic nonlinear Schr\"{o}dinger (NLS) equation in one dimension is one of the basic models of nonlinear optics, photonics, physics of plasma, and hydrodynamics \cite{Fibich,Kev-Dark-2015}. The NLS equation can be modified by the inclusion of the intensity-dependent dispersion in the form 
\begin{equation}
\label{NLS-IDD}
i \partial_t \psi + d(|\psi|^2) \partial_x^2 \psi + \gamma |\psi|^2 \psi = 0,
\end{equation}
where $\gamma$ is the coefficient of the Kerr nonlinearity, $d : (0,\infty) \to \mathbb{R}$ is the intensity-dependent dispersion coefficient, and $\psi = \psi(t,x)$ is the wave function in $(t,x) \in \R \times \R$. If $d(|\psi|^2) = 1$, then the cubic NLS equation (\ref{NLS-IDD}) is focusing for $\gamma > 0$ and defocusing for $\gamma < 0$. It admits bright solitons at the zero background in the former case and dark solitons at the nonzero background in the latter case. 

The NLS equation with nonconstant $d(|\psi|^2)$ has been used in physics literature to model the  coherently prepared multistate atoms \cite{greentree}, quantum well waveguides \cite{koser}, and fiber-optics communication systems \cite{OL2020}, as well as quantum harmonic oscillators in the presence of
nonlinear effective masses~\cite{chang}. The dispersion coefficient $d(|\psi|^2)$ may both decrease and increase with respect to the light intensity \cite{greentree}. Both cases can be modeled in a prototypical form through the dependence $d(|\psi|^2) = 1 - b |\psi|^2$ with $b$ being a constant parameter of either $b > 0$ or $b < 0$ \cite{OL2020}.

The mathematical study of the intensity-dependent NLS models started with \cite{RKP}, where we addressed the model (\ref{NLS-IDD}) with 
$d(|\psi|^2) = 1 - b |\psi|^2$ and $\gamma = 0$. We proved that no bright solitary waves exist for $b < 0$ and a continuous family of  bright solitary waves with the singular profiles exist for $b > 0$. The continuous family can be parameterized by the distance between the two singularities where the wave profile is bounded and the derivative is unbounded. Energetic stability of the entire family of singular solitary waves was proven in \cite{PRK} by using minimization of the mass functional at fixed energy and fixed distance between the two singularities. The stability was obtained for perturbations to the soliton profile in Sobolev space $H^1(\R)$ within a weak formulation where the distance between the two singularities is kept fixed.

Another relevant study was done in \cite{PP24}, where the dark solitary waves were obtained in the case $d(|\psi|^2) = 1/(1-|\psi|^2)$ and $\gamma = 0$. The profiles of dark solitary waves are smooth but the time evolution of the NLS equation is singular. In the particular case of the black solitary waves, it was shown in \cite{PP24} that the stability spectrum of the black solitons consisted of isolated eigenvalues and no continuous spectrum. Similar studies of stability of bright and dark solitary waves were performed in \cite{Albert24,PP23} for the regularized NLS equation proposed earlier in \cite{Lannes} and studied in \cite{AAS}. It was suggested in \cite{PP23} that one can combine the intensity-dependent dispersion term from \cite{PP24} and the regularization term from \cite{Lannes} into the unifying model given by the modified NLS equation. 

Here we consider a different unifying model, where the intensity-dependent dispersion term from \cite{RKP} is combined with the Kerr cubic focusing term. In other words, we address the NLS-IDD equation in the form:
\begin{align}
i \partial_t \psi + (1-|\psi|^2) \partial_x^2 \psi + \gamma |\psi|^2 \psi = 0,
\label{eq:main}
\end{align}
with $\gamma > 0$. Such a model is intrinsically of interest in its own right as a setting presenting the competition of the above mentioned IDD terms
with the standard cubic nonlinearity of relevance to optical
and atomic systems~\cite{kivshar,stringari}. Additionally, 
it is also of interest as a continuum limit of the well-known Salerno model first proposed in~\cite{salerno}. The latter has been used extensively
for exploring the breaking of integrability, the evolution
of conserved quantities, the dynamics of solitary waves, among
many other topics~\cite{cai,mithun}. The Salerno model is written as the lattice 
differential equation 
\begin{equation}
i \partial_t\psi_n + (1-|\psi_n|^2) (\psi_{n+1} + \psi_{n-1}) + \mu |\psi_n|^2 \psi_n = 0,
\label{Salerno}
\end{equation}
where $\mu \in \R$ is the coefficient of the onsite nonlinearity,
$\psi_n = \psi_n(t)$ is the wave function in $(t,n) \in \R \times \Z$, and we have normalized the coefficient of the intersite nonlinearity to unity. 
In the continuum limit where $\psi_n(t) = e^{2it} \psi(h^2 t, hn)$ with small stepsize $h$ 
and smooth $\psi = \psi(t,x)$, we can pick $\mu = 2 + h^2 \gamma$ and obtain (\ref{eq:main}) from (\ref{Salerno}) at the truncated order of $\mathcal{O}(h^2)$. Hence, the NLS-IDD equation also describes the continuum dynamics of the Salerno model with the competing onside and intersite nonlinearity. 

Similarly to other NLS equations with the IDD terms, the NLS-IDD equation (\ref{eq:main}) is a Hamiltonian system with three basic conserved quantitites for the energy 
\begin{align}
\label{energy}
H(\psi) = \int_\R |\partial_x \psi|^2 + \gamma |\psi|^2 + \gamma \log(1 - |\psi|^2) \, dx,
\end{align}
mass
\begin{align}
\label{mass}
Q(\psi) = -\int_\R \log( 1 - |\psi|^2) \, dx,
\end{align}
and momentum 
\begin{align}
\label{momentum}
P(\psi) = i \int_{\R} \frac{\bar{\psi} \partial_x \psi - \psi \partial_x \bar{\psi}}{|\psi|^2} dx,
\end{align}
where $H(\psi)$ and $Q(\psi)$ are well defined in the set of functions in 
\begin{equation}
\label{function-space}
\mathcal{X} := \left\{ \psi \in H^1(\R) : \quad \| \psi \|_{L^{\infty}} < 1 \right\}
\end{equation}
and $P(\psi)$ is well defined for any solution for which $\psi(x) \neq 0$ on $x \in \mathbb{R}$. 

The energy, mass, and momentum are conserved in the time evolution of the NLS-IDD equation (\ref{eq:main}) due to the basic symmetries of translations given by 
\begin{equation}
\label{nls-sym}
\psi(t,x) \mapsto \psi(t+t_0,x+x_0) e^{i \theta_0}, \quad t_0,x_0,\theta_0 \in \mathbb{R}.
\end{equation}
Conservation of $E(\psi)$ in (\ref{energy}) follows from writing (\ref{eq:main}) in the Hamiltonian form 
\begin{align*}
i \partial_t \psi = (1-|\psi|^2) \frac{\delta E}{\delta \bar{\psi}}, \qquad 
\frac{\delta E}{\delta \bar{\psi}} = -\partial_x^2 \psi - \frac{\gamma |\psi|^2 \psi}{1-|\psi|^2},
\end{align*}
from which we obtain 
\begin{align*}
\frac{d}{dt} E(\psi) &= \int_{\R} \left( \frac{\delta E}{\delta \bar{\psi}} \partial_t \psi + \frac{\delta E}{\delta \bar{\psi}} \partial_t \bar{\psi} \right) dx  = 0.
\end{align*}
Conservation of $Q(\psi)$ in (\ref{mass}) follows from the following balance equation obtained from (\ref{eq:main}):
\begin{align*}
i \partial_t \log(1-|\psi|^2) =\partial_x \left( \bar{\psi} \partial_x \psi - 
\psi \partial_x \bar{\psi} \right).
\end{align*}
Conservation of $P(\psi)$ in (\ref{momentum}) can be checked directly as 
\begin{align*}
\frac{d}{dt} P(\psi) &= \int_{\R} \left( \frac{\partial_x \psi \partial_x^2 \psi}{\psi^2} - \frac{\partial_x^3 \psi}{\psi} +  \frac{\partial_x \bar{\psi} \partial_x^2 \bar{\psi}}{\bar{\psi}^2} - \frac{\partial_x^3 \bar{\psi}}{\bar{\psi}} \right) dx \\
& \quad + \int_{\R} \left( \bar{\psi} \partial_x^3 \psi + \partial_x \bar{\psi} \partial_x^2 \psi + \psi \partial_x^3 \bar{\psi} + \partial_x \psi \partial_x^2 \bar{\psi} \right) dx \\
& \quad - 2 \gamma \int_{\R} \partial_x |\psi|^2 dx = 0.
\end{align*}
The momentum $P(\psi)$ corresponds to the renormalized momentum, which is the only momentum-type conserved quantity in the NLS equation with the IDD terms, see \cite{PP24}.

Bright solitons of the NLS-IDD equation (\ref{eq:main}) are the standing wave solutions of the form 
\begin{align*}
\psi(x,t) = e^{i\om t} \vp_\om(x)
\end{align*}
with the frequency $\om>0$ and the profile $\vp_\om$ being a real, spatially decaying solution of the second-order differential equation 
\begin{align}
\label{eq:odephi}
\frac{d^2 \vp}{dx^2} = \frac{(\om-\gamma \vp^2)}{1-\vp^2}\vp = -\frac{dV}{d\vp}
\end{align}
associated with the potential energy
\begin{align}
\label{eq-V}
V(\varphi) := \frac{\om-\gamma}{2}\log|1-\varphi^2| - \frac{\gamma}{2}\varphi^2.
\end{align}
The existence of bright solitons with the smooth profile $\vp_{\om}$ is given by the following theorem. 

\begin{theorem}
	\label{th-existence}
	Fix $\gamma > 0$. There exists the solitary wave solution with $\varphi_{\omega} \in H^{\infty}(\R)$ to the second-order equation (\ref{eq:odephi}) if and only if $\omega \in (0,\gamma)$. 
	Moreover, the family $\{ \varphi_{\omega}\}_{\omega \in (0,\gamma)}$ is also smooth with respect to $\omega$.
\end{theorem}

\begin{remark}
	\label{rem-limits}
In the limit $\omega \to 0$, the size of $\varphi_{\omega}$ is small according to the formal asymptotic expansion
\[
\vp_\om(x) = \varepsilon \phi_{\Omega}(\varepsilon x) + \mathcal{O}(\varepsilon^3), \qquad \om = \varepsilon^2 \Omega,
\]
where the profile $\phi_{\Omega}$ is found from the second-order equation
\[
\phi'' = \Omega \phi - \gamma \phi^3
\]
for every fixed $\Omega > 0$, e.g. in the explicit form 
$$
\phi_{\Omega}(x) = \frac{\sqrt{2\Omega}}{\sqrt{\gamma}} {\rm sech}(\sqrt{\Omega} x).
$$ 
In the limit $\omega \to \gamma$, the second-order equation (\ref{eq:odephi}) becomes linear, e.g.  
$\vp'' = \gamma \vp$, with the formal peakon solution 
$$
\vp_{\gamma}(x) = e^{-\sqrt{\gamma}|x|}.
$$ 
The case $\omega > \gamma$ gives a family of bright solitons with the singular profiles and the singularity is the same as  the one considered in \cite{PRK,RKP} for $\gamma = 0$. 
\end{remark}

The main motivation for our work is to establish the energetic stability of the bright solitons with the smooth profile $\vp_{\omega}$ for $\omega \in (0,\gamma)$ under the presence of the intensity--dependent dispersion. The following theorem presents the main result.

\begin{theorem}
	\label{theorem-main}
	Let $\vp_{\omega} \in H^{\infty}(\R)$ be the spatial profile satisfying (\ref{eq:odephi}) for $\omega \in (0,\gamma)$, according to Theorem \ref{th-existence}. Then, it is a local nondegenerate (up to two symmetries) minimizer of the augmented energy $\Lambda_{\omega} := H + \omega Q$ subject to fixed mass $Q$ in $H^1(\R)$ if and only if the mapping $\omega \mapsto Q(\vp_{\omega})$ is monotonically increasing. 
\end{theorem}

\begin{remark}
	If local well-posedness of the NLS-IDD equation (\ref{eq:main}) can be established in the function set $\mathcal{X}$ in (\ref{function-space}), then Theorem \ref{theorem-main} yields the orbital stability of bright solitons along the orbit $\{ \vp_{\omega}(\cdot - \xi) e^{i \theta}\}_{\xi,\theta \in \R}$ in $H^1(\R)$. However, the quasilinear NLS equation of the class (\ref{eq:main}) is locally well-posed in $H^{\infty}(\R)$, see \cite{Feola,KPV03,Popen}, whereas the result of Theorem \ref{theorem-main} allows us to control perturbations to the bright soliton profile in the $H^1(\R)$ norm only. 
\end{remark}

\begin{remark}
	\label{remark-stab}
We show numerically that there exist $\omega_1, \omega_2$ satisfying
$0 < \omega_1 < \omega_2 < \gamma$ such that 
the mapping $\omega \mapsto Q(\vp_{\omega})$ is monotonically increasing if 
$\omega \in (0,\omega_1) \cup (\omega_2,\gamma)$ and 
monotonically decreasing if $\omega \in (\omega_1,\omega_2)$. 
The former is energetically stable and the latter is energetically unstable. 
Stability of smooth solitary waves in the limits $\omega \to 0$ and $\omega \to \gamma$ agree with the stability of bright solitons in the cubic NLS equation and the energetic stability of singular solitary waves in the NLS-IDD equation for $\gamma = 0$ proven in \cite{PRK}.
\end{remark}

\begin{remark}
The  spectral instability of solitary waves for
$\omega \in (\omega_1,\omega_2)$ has also been observed in various
models involving a modification of the standard dispersion
and cubic nonlinearity. Such examples involve the discrete NLS equations with a long-range dispersion~\cite{gaididei}, as well
as the generalized NLS equation where the nonlinearity term
features a non-cubic (higher order) power~\cite{malwei}. 
\end{remark}

The paper is organized as follows. Existence of bright solitons with smooth profiles is considered in Section \ref{sec-2} with the phase plane analysis, where the proof of Theorem \ref{th-existence} is given. Stability of bright solitons with the proof of Theorem \ref{theorem-main} is developed in Section \ref{sec-3} with analysis of the Hessian operator for the augmented energy $\Lambda_{\omega} = H + \omega Q$. Numerical results are described in Section \ref{sec-4}, where we approximate the mapping $\omega \mapsto Q(\varphi_{\omega})$, eigenvalues of the spectral stability problem, and  the time-dependent evolution of the NLS equation \eqref{eq:main} suggesting that spectrally stable bright solitons are also dynamically (nonlinearly) stable. An outlook of open directions of study is given in Section \ref{sec-5}.

\section{Existence of bright solitons with smooth profiles}
\label{sec-2}

Here we fix $\gamma > 0$ and consider solutions of the second-order equation (\ref{eq:odephi}) on the phase plane $(\varphi,\varphi') \in \mathbb{R}^2$.
First we consider the case $\omega > 0$, for which there exist three local extrema of $V$ at $0$ and $\pm \sqrt{\om/\gm}$ as well as two singularities at $\pm 1$, see  (\ref{eq-V}). For convenience, we denote $\varphi_* :=\sqrt{\om/\gm}$. Since 
\begin{align*}
V''(\varphi) = -\frac{\om(1+\varphi^2) + \gm \varphi^2(\varphi^2-3)}{(1-\varphi^2)^2}.
\end{align*}
we obtain 
\begin{align*}
V''(\varphi_*) = \frac{2\omega \gamma}{\omega - \gamma}.
\end{align*}
Hence $\pm \varphi_*$ are minima of $V$ for $\om \in (0,\gm)$ and 
maxima of $V$ for $\omega \in (\gm,\infty)$, whereas $0$ is always a maximum of $V$ if $\omega > 0$. Different cases are considered next.

\begin{enumerate}
\item \underline{$\om \in (0,\gm)$.} Since $\varphi_* \in (0,1)$, there are three equilibrium points in the vertical strip $[-1,1] \times \mathbb{R}$ for 
$(\varphi,\varphi')$. The origin $(0,0)$ is a saddle point and 
$(\pm \varphi_*,0)$ are the center points since $V$ is similar to a double-well potential in $[-1,1]$ with $V(\varphi) \to +\infty$ as $\varphi \to \pm 1$. The phase portrait of the planar Hamiltonian system described by the second-order equation (\ref{eq:odephi}) is shown on Figure \ref{fig-1} from the level curves of the function 
\begin{align*}
E(\varphi,\varphi') := \frac{1}{2} (\varphi')^2 + V(\varphi),
\end{align*}
which is $x$-independent on every smooth solution of (\ref{eq:odephi}).
Periodic orbits exist in the vertical strip $[-1,1] \times \mathbb{R}$ for every $E \in (V(\varphi_*),0) \cup (0,\infty)$ either inside or outside the homoclinic orbits for $E = 0$. The homoclinic orbits with $E=0$, correspond to the bright soliton with the smooth profile $\varphi_{\omega} \in H^{\infty}(\R)$ of Theorem \ref{th-existence}. Its maximum value is the unique zero $\varphi_0 \in (0,1)$ of $V(\varphi) = 0$. All orbits outside $[-1,1] \times \mathbb{R}$ diverge to infinity and do not give the solitary wave profiles decaying to zero at infinity. Since $V$ is smooth in $\omega$ and $\varphi_{\omega}$ is bounded away from $1$ for every $\omega \in (0,\gamma)$, the spatial profile $\varphi_{\omega}$ is smooth with respect to $\omega$. Smooth profiles $\varphi_{\omega}$ are shown on Figure \ref{fig:vps_x}. The slopes grow as the values of $\omega$ increase towards the value of $\gamma = 1$. \\

\begin{figure}[htb!]
	\includegraphics[width=12cm,height=10cm]{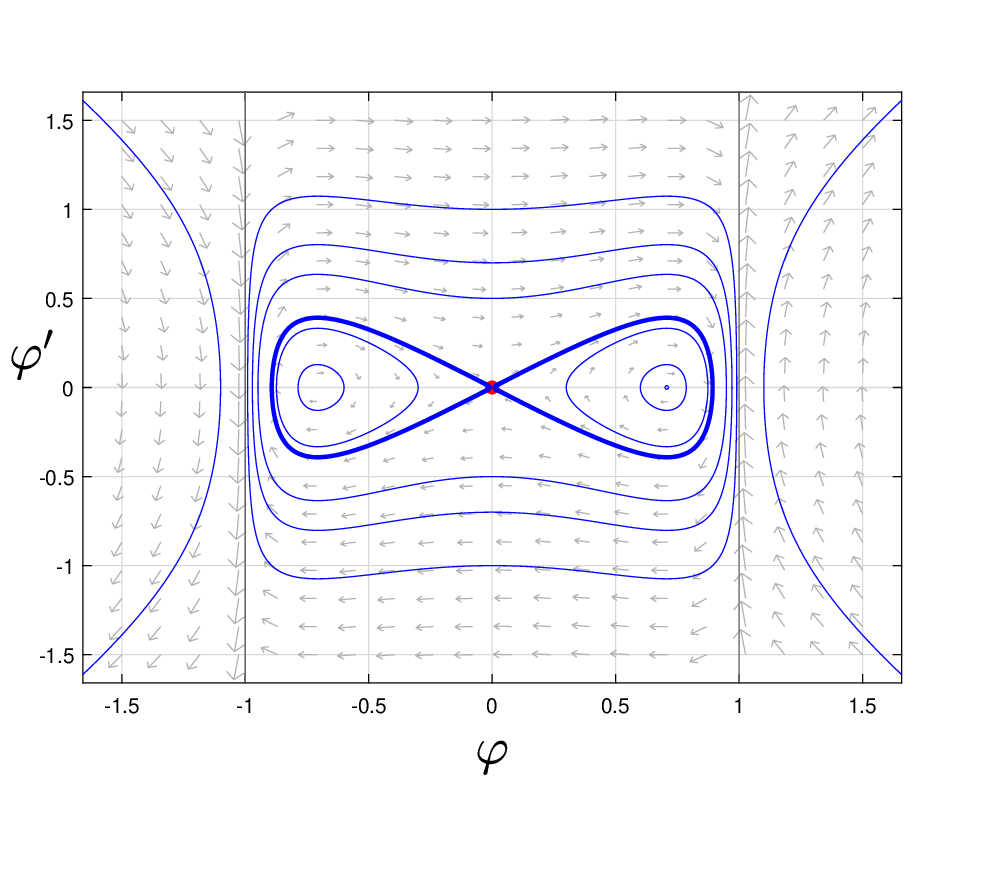}
	\caption{Phase portrait for the second-order equation (\ref{eq:odephi}) with $\om \in (0,\gm)$ for $\gamma = 1$. The two homoclinic orbits correspond to the smooth profiles $\pm \vp_\om$ with values in $(-1,1)$.}
	\label{fig-1}
\end{figure}

\begin{figure}[htb!]
	\includegraphics[width=10cm,height=6cm]{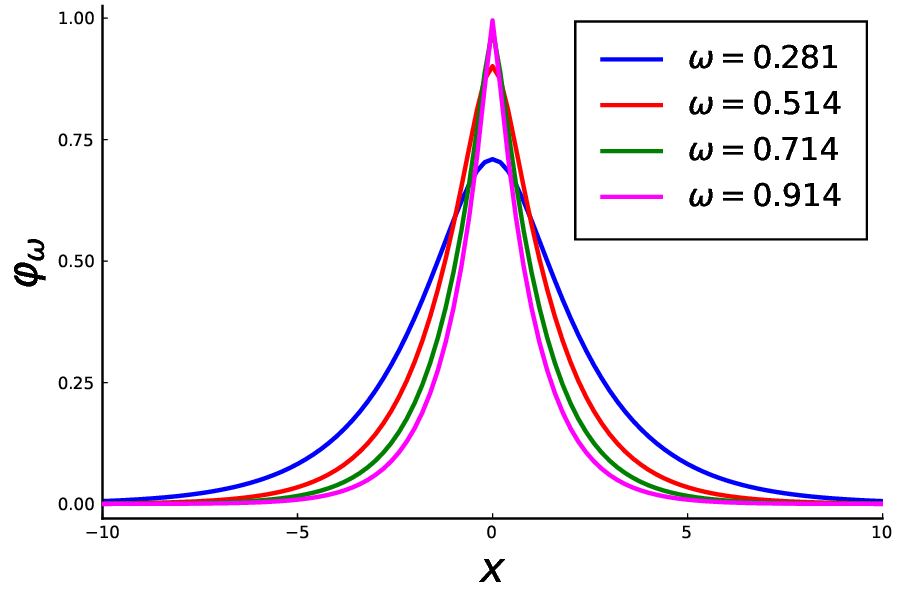}
	\caption{Bright soliton profiles $\vp_\om$ for various values of the frequency $\om$.}
	\label{fig:vps_x}
\end{figure}

\item \underline{$\om=\gm$.} The singularity is canceled for 
$V(\varphi) = -\frac{1}{2} \varphi^2$ and all smooth solutions 
of the linear equation $\varphi'' = \gamma \varphi$ diverge to infinity. See Remark \ref{rem-limits} for a peakon solution with the 
profile $\varphi_{\gamma} \in H^1(\R)$. Existence of such weakly singular bright soliton is beyond the scope of this work. \\

\item \underline{$\om \in (\gm,\infty)$.} Since $\pm \varphi_*$ are now maxima of $V$ outside $[-1,1]$, we have $V(\varphi) \to -\infty$ as $\varphi \to \pm 1$. The other maximum is at $0$ with $V(0) = 0$, so that we compute
\begin{align*}
V_*(\gamma,\om) := V(\varphi_*) = \frac{\om-\gm}{2}\log\left(\frac{\om}{\gm}-1\right) - \frac{\om}{2}, \quad \text{ for } \om \in (\gm,\infty).
\end{align*}
A simple analysis of this function on Figure \ref{fig:Vstar} shows that for each $\gm>0$ fixed, there is a unique $\om=\om_*(\gm) \in (\gm,\infty)$ such that $V_*(\gm,\om_*(\gm))=0$ and $V_*(\gm,\om) \lessgtr 0$ for $\om \lessgtr \om_*(\gm)$. 

\begin{figure}[htb!]
	\includegraphics[width=8cm,height=6cm]{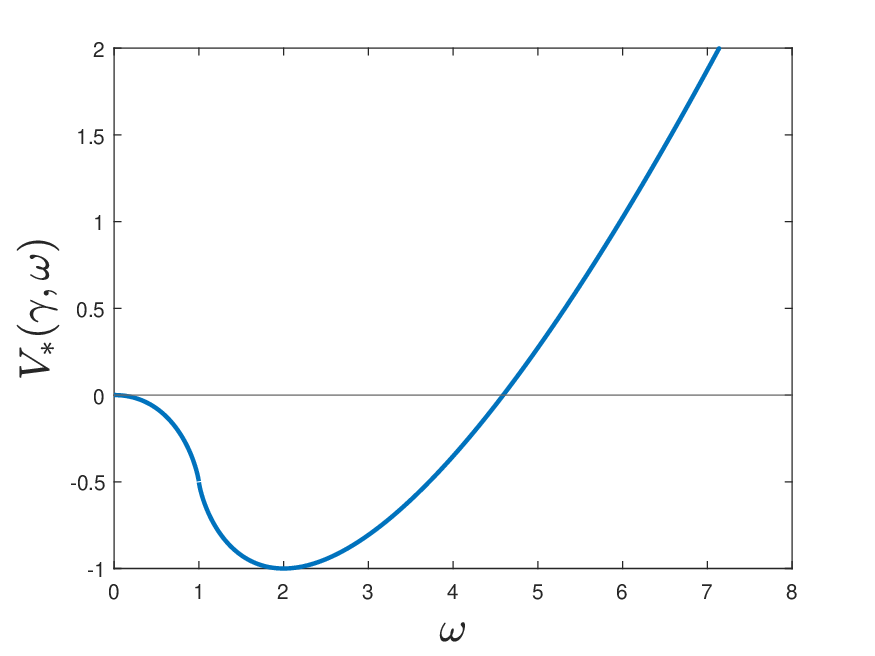}
	\caption{The function $V_*(\gm,\om)$ vs. $\om$ for fixed $\gm=1$. The unique zero of $V_*$ occurs at $\om=\om_*(\gm)$.}
	\label{fig:Vstar}
\end{figure}

Figure \ref{fig-2} shows the phase portraits of the second-order equation (\ref{eq:odephi}) for the two cases: one is for  $\om \in (\gm,\om_*(\gm))$ (left) and the other one is for $\om \in (\om_*(\gm),\infty)$ (right). 
In both cases, all smooth solutions 
of the second-order equation (\ref{eq:odephi}) diverge to infinity. The homoclinic orbits for the level $E = 0$ are broken at the singularities $\pm 1$. The smooth orbits outside the vertical strip $[-1,1] \times \mathbb{R}$ 
for the same level $E = 0$ either diverge to infinity for $\om \in (\gm,\om_*(\gm))$ or are bounced back to the singularities $\pm 1$ 
for $\om \in (\om_*(\gm),\infty)$. The latter case resembles the case of $\gamma = 0$, for which we constructed 
a continuous family of bell-shaped solitary waves with the singular profiles in \cite{RKP} 
within a weak formulation of the second-order equation (\ref{eq:odephi}) for $\gamma = 0$, see also \cite{PRK}.
%Existence of such singular bright solitons is again beyond the scopes of this work. 
Per the consideration of the latter work, we do not examine such
waveforms further herein.

\begin{figure}[htb!]
	\includegraphics[width=8cm,height=6cm]{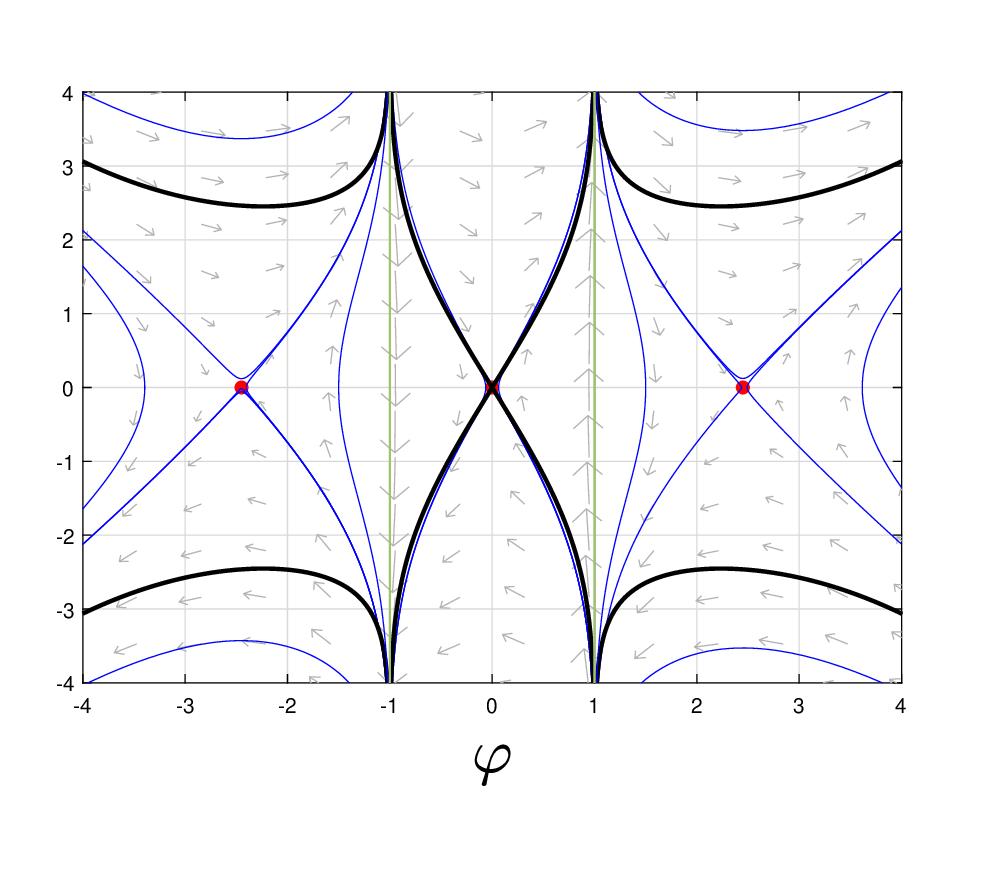}
	\includegraphics[width=8cm,height=6cm]{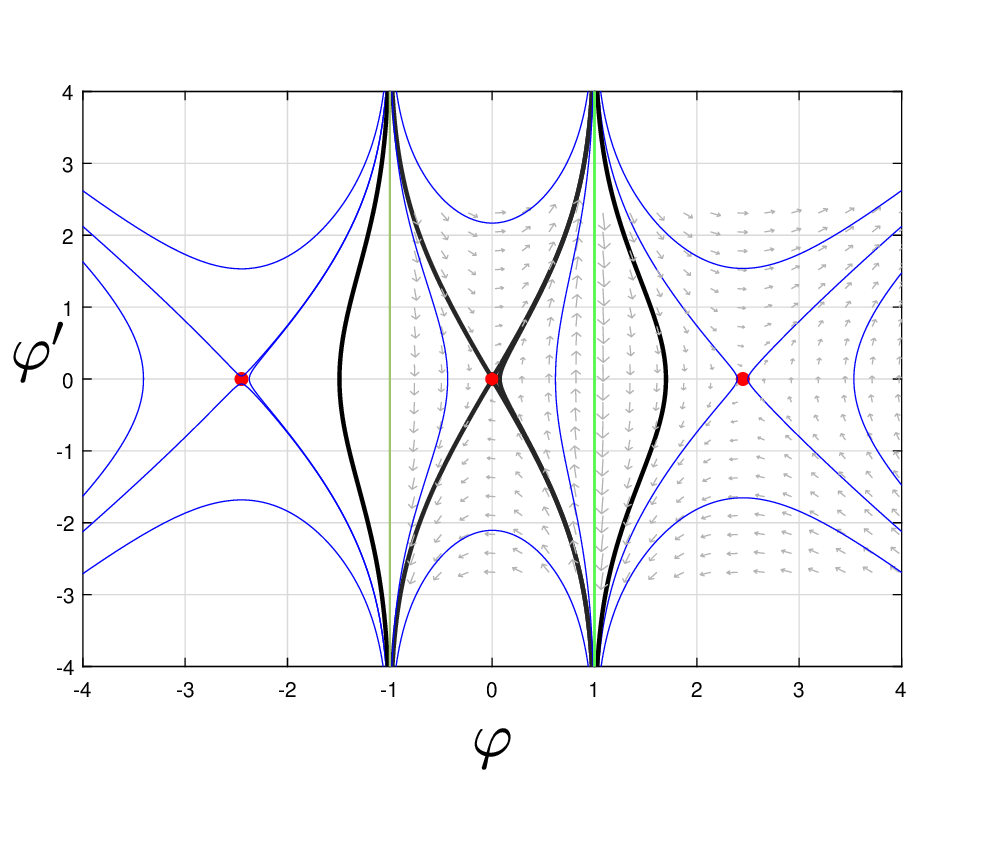}
	\caption{Phase portraits for the second-order equation \eqref{eq:odephi} with $\gm=1$ for which $\om_*(1)\approx 4.5$. We use $\om=3$ for $\om \in (\gm,\om_*(\gm))$ (left), and $\om = 6$ for $\om \in (\om_*(\gm),\infty)$ (right). In both panels, the green vertical lines indicate the singularities at $\vp = \pm 1$, and the black curves are the level curves $E(\vp,\vp') = 0$.}
	\label{fig-2}
\end{figure}

\end{enumerate}

For $\omega \leq 0$, there is only one local extremum of $V$ at $0$ 
and it is a local minimum of $V$ with $V(\varphi) \to +\infty$ as $\varphi \to \pm 1$. There is a continuous family of smooth periodic orbits of the second-order equation (\ref{eq:odephi}) in the vertical strip $[-1,1] \times \mathbb{R}$ and a continuous family of unbounded orbits outside the strip.
There exist no homoclinic orbits for the bright solitons. 

Based on the above analysis, the proof of Theorem \ref{th-existence} is complete.

\section{Stability of bright solitons with smooth profiles}
\label{sec-3}

To simplify the proof of Theorem \ref{theorem-main}, here we set $\gamma = 1$. 
This can be considered without loss of generality since $\gamma$ can be scaled to unity by a scaling transformation of the  $(x,t)$ variables in the NLS-IDD equation (\ref{NLS-IDD}). 

Let $\varphi_{\omega} \in H^{\infty}(\R)$ be the spatial profile satisfying (\ref{eq:odephi}) for $\omega \in (0,1)$ with $\gamma = 1$, according to Theorem \ref{th-existence}. Adding a perturbation $u+iv$ to the profile $\varphi_{\omega}$ and linearizing yields the spectral stability problem 
in the form:
\begin{align}
\label{eq:NLS_stab}
\begin{pmatrix}
0 & \mL_- \\
-\mL_+ & 0
\end{pmatrix}
\begin{pmatrix}
u \\ v
\end{pmatrix}
&= \lm
\begin{pmatrix}
u \\ v
\end{pmatrix}
\end{align}
where
\begin{align*}
\mL_- &:= -(1-\vp_\om^2)\partial_x^2 + \om - \vp_\om^2,
\\
\mL_+ &:= -(1-\vp_\om^2)\partial_x^2 + \om + 2\vp_\om \partial_x^2 \vp_\om - 3 \vp_\om^2. 
\end{align*}
Since the coefficient $(1-\varphi_\om^2)$ is sign-definite for the profile $\vp_\om \in H^{\infty}(\R)$ of Theorem \ref{th-existence}, 
the weight $(1-\vp_\om^2)^{-1}$ is bounded away from $0$ and converges 
to $1$ as $|x| \to \infty$. The linear Schr\"{o}dinger operators
\begin{equation}
\label{Hessian}
\mathcal{S}_{\pm} := (1-\varphi_{\omega}^2)^{-1} \mathcal{L}_{\pm} : H^2(\R) \subset L^2(\R) \to L^2(\R)
\end{equation}
are self-adjoint so that we can consider the spectral stability problem (\ref{eq:NLS_stab}) in the weighted Hilbert space $\mc H \times \mc H$, where $\mc H := L^2(\R,(1-\vp_\om^2)^{-1} \, dx)$ is equipped with the inner product 
\begin{equation*}
\langle \cdot,\cdot \rangle_{\mc H} := \langle (1-\vp_\om^2)^{-1} \cdot,\cdot \rangle_{L^2}.
\end{equation*}
This approach is very similar to the study of stability in the regularized NLS equation, see \cite{PP24}. Since $(1-\varphi_{\pm}^2)$ is bounded away from $0$, we can reformulate the spectral stability problem (\ref{eq:NLS_stab}) in the equivalent form 
\begin{align}
\label{eq:NLS_stab-equiv}
\begin{pmatrix}
0 & \mathcal{M}_- \\
-\mathcal{M}_+ & 0
\end{pmatrix}
\begin{pmatrix}
\tilde{u} \\ \tilde{v}
\end{pmatrix}
&= \lm
\begin{pmatrix}
\tilde{u} \\ \tilde{v}
\end{pmatrix},
\end{align}
where $(\tilde{u},\tilde{v}) = (1-\varphi_{\omega}^2)^{-1/2} (u,v)$ 
and 
\begin{equation}
\label{operators}
\mathcal{M}_{\pm} := (1-\varphi_{\omega}^2)^{1/2} \mathcal{S}_{\pm} (1-\varphi_{\omega}^2)^{1/2} : H^2(\R) \subset L^2(\R) \to L^2(\R)
\end{equation}
are also self-adjoint operators. 

For the proof of the energetic stability of bright solitons with the smooth profile $\varphi_{\omega}$, we follow the standard algorithm of placing $\varphi_{\omega}$ in the variational context as a critical point 
of the augmented energy $\Lambda_{\omega} := H + \omega Q$, where $H$ and $Q$ are given by (\ref{energy}) and (\ref{mass}). Then, we show that 
$\mathcal{S}_{\pm}$ in (\ref{Hessian}) are Hessian operators 
of the variational problem and that their combined spectra 
in $L^2(\R)$ includes a simple negative eigenvalue, a double zero 
eigenvalue, and a strictly positive part bounded away from zero. 
Finally, we show that $\varphi_{\omega}$ is a local nondegenerate (up to two symmetries) minimizer of $\Lambda_{\omega}$ subject to fixed mass $Q$ if and only if the mapping $\omega \mapsto Q(\varphi_{\omega})$ is monotonically increasing. This yields the criterion for energetic stability of the bright solitons with the smooth profile $\varphi_{\omega}$ given by Theorem \ref{theorem-main}.

For a linear operator $T: \mc D(T) \subset \mc H \to \mc H$ with a dense domain $\mc D(T)$ in a Hilbert space $\mc H$, we denote 
\begin{align*}
 n(T) = \dim \{v \in \mc H: \quad \langle Tv,v \rangle_{\mc H} < 0\},
\end{align*}
which is the number of negative eigenvalues of $T$ with the account of their multiplicity. Similarly, we denote multiplicity of the zero eigenvalue of $T$ by $z(T)$. The algorithm of the proof of Theorem \ref{theorem-main} is divided into several steps. 

\vspace{0.25cm}

\underline{\rm Step 1:} $\varphi_{\omega}$ is a solution of the Euler--Lagrange equation for $\Lambda_{\omega} = H + \omega Q$.

 \vspace{0.25cm}
 
 To prove Step 1, we compute variational derivative of $H$ and $Q$ with respect to $\bar{\psi}$:
 $$
 \frac{\delta H}{\delta \bar{\psi}} = -\partial_x^2 \psi - \frac{\gamma |\psi|^2 \psi}{1 - |\psi|^2}, \qquad  \frac{\delta Q}{\delta \bar{\psi}} = \frac{\psi}{1-|\psi|^2},
 $$
 so that the second-order equation (\ref{eq:odephi}) is written as 
 $$
  \frac{\delta H}{\delta \bar{\psi}} + \omega  \frac{\delta Q}{\delta \bar{\psi}} = 0.
 $$
 Hence, $\varphi_{\omega}$ is a critical point of $\Lambda_{\omega}$.
 
 \vspace{0.25cm}
 
 \underline{\rm Step 2:} Operators $\mathcal{S}_{\pm}$ in (\ref{Hessian}) are Hessian operators of $\Lambda_{\omega}$ at $\varphi_{\omega}$.
 
 \vspace{0.25cm}
 
 To prove Step 2, we add perturbation $u+iv$ to $\varphi_{\omega}$, use Step 1, and derive
 $$
\Lambda_{\omega}(\varphi_{\omega} + u + iv) - \Lambda_{\omega}(\varphi_{\omega}) = \langle \mathcal{S}_+ u, u \rangle_{L^2} +  \langle \mathcal{S}_- v, v \rangle_{L^2} + \mathcal{O}(\| u + iv \|_{H^1}^3),
 $$
 where $\mathcal{S}_{\pm}$ are given by (\ref{Hessian}). 
 Hence, $\mathcal{S}_{\pm}$ are Hessian operators of $\Lambda_{\omega}$ at $\varphi_{\omega}$.

 \vspace{0.25cm}
 
 \underline{\rm Step 3:} $n(\mathcal{S}_-) = 0$ and $z(\mathcal{S}_-) = 1$ 
 in $L^2(\R)$.
 
 \vspace{0.25cm}
 
 To prove Step 3, we recall that $\mathcal{S}_- = (1-\varphi_{\omega}^2)^{-1} \mathcal{L}_-$. Due to phase rotation symmetry, we have $\mc L_- \vp_\om = 0$ with $\vp_\om(x) > 0$ and $\vp_\om \in H^2(\R)$. The essential spectrum of  $\mathcal{S}_-$ in $L^2(\R)$ is given by $[\omega,\infty)$ by Weyl's theorem. 
Since it is bounded from zero by a positive constant $\omega > 0$, positivity of 
$\varphi_{\omega}$ implies by Sturm's Theorem that $n(\mc S_-) = 0$ and $z(\mc S_-) = 1$.

 \vspace{0.25cm}

\underline{\rm Step 4:} $n(\mathcal{S}_+) = 1$ and $z(\mathcal{S}_+) = 1$ 
in $L^2(\R)$.

\vspace{0.25cm}

To prove Step 4, we recall that $\mathcal{S}_+ = (1-\varphi_{\omega}^2)^{-1} \mathcal{L}_+$. Due to spatial translation symmetry, we have $\mc L_+ \partial_x \vp_\om = 0$ with $\partial_x \vp_\om \in H^2(\R)$ having exactly one zero on $\R$. 
Since 
\begin{equation}
\label{tech-eq}
\omega + 2\vp_\om \partial_x^2 \vp_\om - 3 \vp_\om^2 = 
\omega \frac{1 + \varphi_{\omega}^2}{1 - \varphi_{\omega}^2} -  
\varphi_{\omega}^2 \frac{3- \varphi_{\omega}^2}{1 - \varphi_{\omega}^2},
\end{equation}
and $\varphi_{\omega}(x) \to 0$ as $|x| \to \infty$ exponentially fast, 
the essential spectrum of  $\mathcal{S}_+$ in $L^2(\R)$ is given by $[\omega,\infty)$ by Weyl's theorem. 
Since it is bounded from zero by a positive constant $\omega > 0$, a single zero of $\partial_x \varphi_{\omega}$ on $\R$ implies by Sturm's Theorem that $n(\mc S_-) = 1$ and $z(\mc S_-) = 1$.

\vspace{0.25cm}

\underline{\rm Step 5:} $\varphi_{\omega}$ is a local nondegenerate (up to two symmetries) minimizer of $\Lambda_{\omega}$ subject to fixed mass $Q$ if and only if the mapping $\omega \mapsto Q(\varphi_{\omega})$ is monotonically increasing.

\vspace{0.25cm}

To prove Step 5, we need to show $n(\mathcal{S}_+ |_{\{ v_{\omega} \}^{\perp}}) = 0$ 
and $z(\mathcal{S}_+ |_{\{ v_{\omega} \}^{\perp}}) = 1$, where 
$$
v_{\omega} := \frac{\varphi_{\omega}}{1 - \varphi_{\omega}^2}
$$
represents the first variation of $Q$ at $\varphi_{\omega}$ as in Step 1 and $\mathcal{S}_+ |_{\{ v_{\omega} \}^{\perp}}$ denotes the restriction of $\mathcal{S}_+$ to the constrained $L^2(\R)$ space with a scalar orthogonality condition $\langle \cdot, v_{\omega} \rangle_{L^2} = 0$. 
As is well-known, if $n(\mathcal{S}_+) = 1$, $n(\mathcal{S}_-) = 0$, and $z(\mathcal{S}_{\pm}) = 1$ as in Steps 3 and 4 and $\mathcal{S}_{\pm}$ are Hessian operators for $\Lambda_{\omega}$ at $\varphi_{\omega}$ as in Step 2, 
then the assertion is true if and only if 
$$
\langle \mathcal{S}^{-1}_+ v_{\omega}, v_{\omega} \rangle_{L^2} < 0.
$$
Recall by Theorem \ref{th-existence} that $\varphi_{\omega} \in H^{\infty}(\R)$ is also smooth with respect to $\omega$ for $\omega \in (0,1)$. By differentiating of equation (\ref{eq:odephi}) in $\omega$ for $\omega \in (0,1)$
and comparing with (\ref{tech-eq}), we get 
$$
\mathcal{S}_+ \partial_{\omega} \varphi_{\omega} = -v_{\omega}.
$$
Hence 
$$
\langle \mathcal{S}^{-1}_+ v_{\omega}, v_{\omega} \rangle_{L^2} = -\langle \partial_{\omega} \varphi_{\omega},v_{\omega} \rangle_{L^2} = -\frac{1}{2} \partial_{\omega} Q(\varphi_{\omega}),
$$
and the assertion is true if and only if the mapping $\omega \mapsto Q(\varphi_{\omega})$ is monotonically increasing.

\vspace{0.25cm}

Based on the above five steps, the proof of Theorem \ref{theorem-main} is complete. 

\begin{remark}
	If the bright soliton with the profile $\varphi_{\omega}$ is a local nondegenerate (up to two symmetries) minimizer of $\Lambda_{\omega}$ subject to fixed mass $Q$ as in Theorem \ref{theorem-main}, then the linear spectral problem 
	\begin{align*}
	\begin{pmatrix}
	0 & \mathcal{S}_- \\
	-\mathcal{S}_+ & 0
	\end{pmatrix}
	\begin{pmatrix}
	u \\ v
	\end{pmatrix}
	&= \lm
	\begin{pmatrix}
	u \\ v
	\end{pmatrix}
	\end{align*}
	admits no eigenvalues $\lambda \in \C \backslash \{ i \R \}$ with eigenvectors $(u,v) \in H^2(\R) \times H^2(\R)$. Since the weight  $(1-\varphi_{\omega}^2)^{-1/2}$ is bounded away from $0$ and $\infty$, 
	Sylvester's inertia law theorem implies the same counts of negative and zero 
	eigenvalues of $\mathcal{M}_{\pm}$ in $L^2$ for $\mathcal{M}_{\pm}$ given by (\ref{operators}). This implies the same stability result 
	in the spectral problem (\ref{eq:NLS_stab-equiv}). Finally, by the same boundedness of $(1-\varphi_{\omega}^2)^{-1/2}$ and the transformation of eigenvector 
	$(u,v) \in H^2(\R) \times H^2(\R)$ to $(\tilde{u},\tilde{v}) \in H^2(\R) \times H^2(\R)$, it implies that 
	$n(\mathcal{L}_+) = 1$, $n(\mathcal{L}_-) = 0$, and $z(\mathcal{L}_{\pm}) = 1$ in $\mc H$ and that the same stability result holds 
	in the spectral problem (\ref{eq:NLS_stab}).
\end{remark}

\section{Numerical results}
\label{sec-4}

Here we illustrate numerically the statements in Theorem \ref{theorem-main} and Remark \ref{remark-stab}, regarding monotonicity changes in the mapping $\om \mapsto Q(\vp_\om)$ and the spectral stability of the bright solitons in the stability problem \eqref{eq:NLS_stab}. In addition, we perform several numerical experiments which suggest that spectrally stable solitons are also dynamically stable. 

Profiles $\varphi_{\omega}$ of bright solitons are obtained numerically in the full range $\om \in (0,\gm)$ for $\gm=1$ via a standard pseudo-arclength continuation, using the continuation package \texttt{BifurcationKit.jl} in Julia. At each continuation step, the profile is obtained by the GMRES method with a diagonal (Jacobi) preconditioner.

\begin{figure}[htb!]
	\includegraphics[scale=0.5]{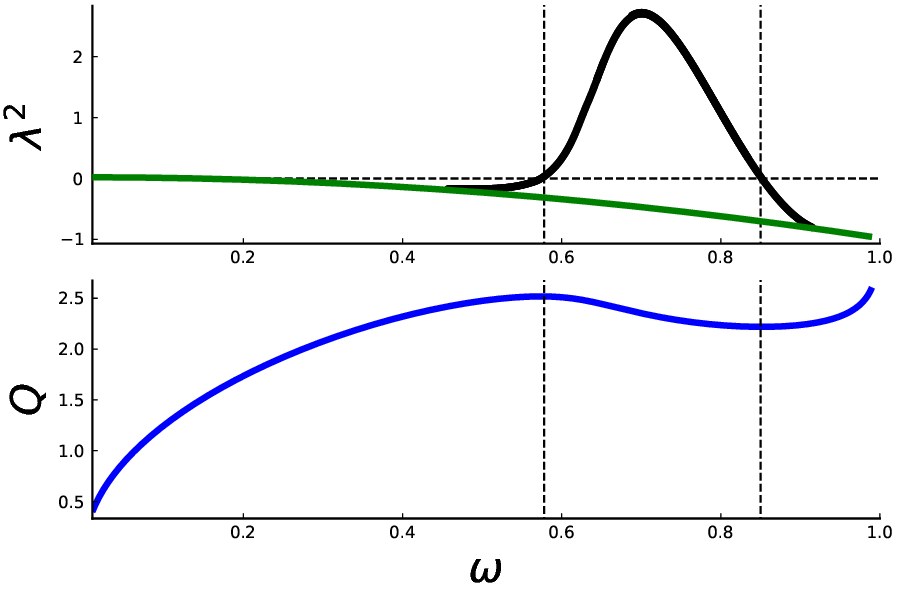}
	\caption{Top: Square of the bifurcating unstable eigenvalue $\lm^2$ for the spectral stability problem \eqref{eq:NLS_stab}. Bottom: the map $\om \mapsto Q(\vp_\om)$. In the top panel, the dashed horizontal line is drawn at $\lm^2 = 0$ and the solid green curve is the function $\om \mapsto -\om^2$, which is the boundary of the continuous spectrum. In both panels, the dashed vertical lines are drawn at $\om_1$ and $\om_2$.}
	\label{fig-lm2-Q}
\end{figure}

According to Theorem \ref{theorem-main}, the bright soliton with the smooth profile $\vp_\om$ is stable if and only if the mapping $\om \mapsto Q(\vp_\om)$ is increasing. Numerical computations show that this function has two extrema at $\om_1, \om_2 \in (0,\gm)$. The former is seen to be a maximum while the latter is a minimum, leading to two switches in the stability of the bright solitons. These are indicated by the two vertical lines in the bottom panel of Figure \ref{fig-lm2-Q}.

\begin{figure}[htb!]
	\includegraphics[scale=0.4]{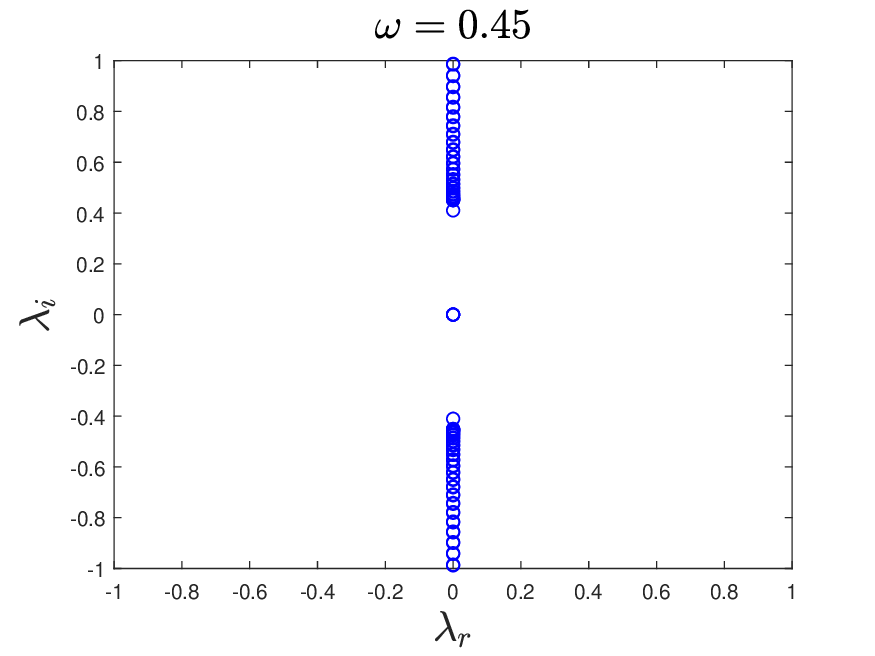}
	\includegraphics[scale=0.4]{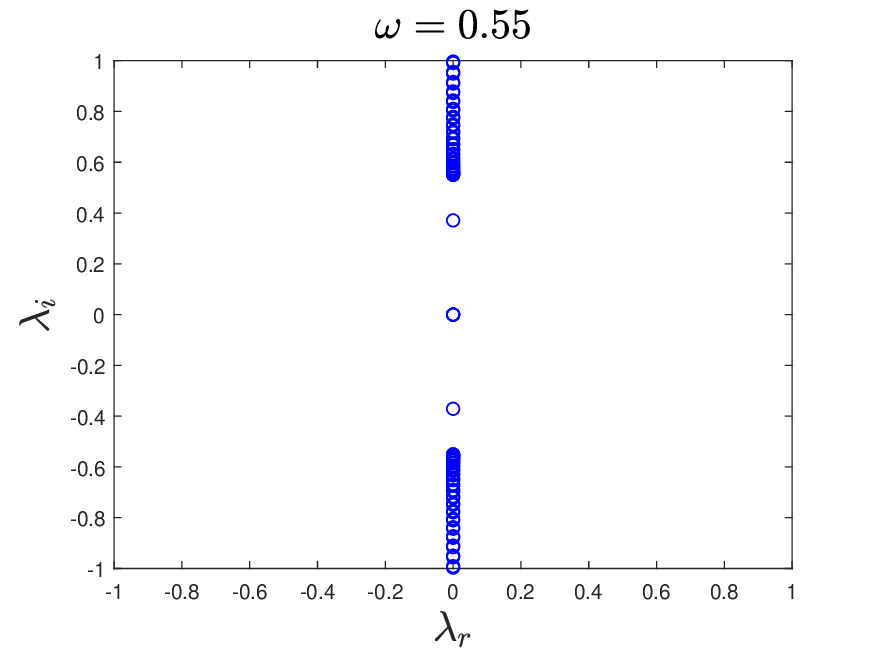}
	\\
	\includegraphics[scale=0.4]{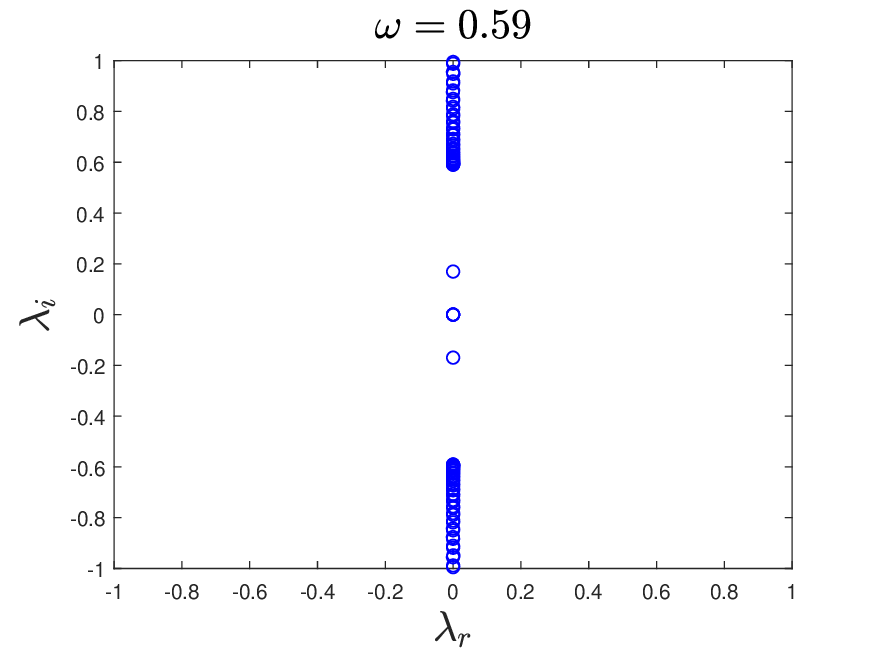}
	\includegraphics[scale=0.4]{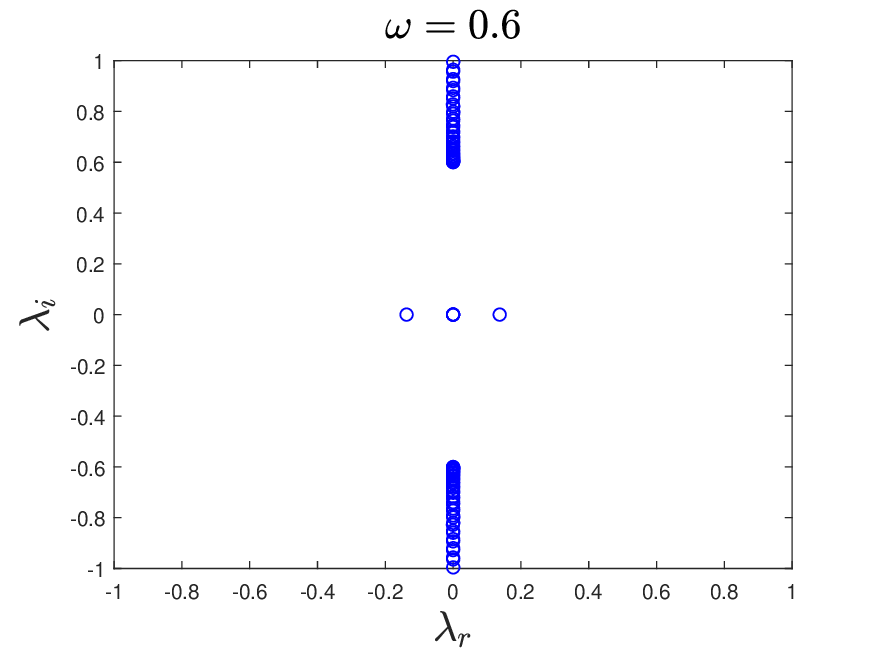}
	
	\caption{Complex plane $(\lm_r,\lm_i)$ for eigenvalues $\lm = \lm_r + i\lm_i$ of the spectral stability problem \eqref{eq:NLS_stab}, for $\om$ values near the first eigenvalue crossing. As $\om$ increases, the nonzero eigenvalue pair moves towards the origin, crossing through onto the real axis around $\om_1 \approx 0.592$. The situation at the second crossing is reversed, with the same eigenvalue pair crossing from the real axis back to the imaginary axis as $\om$ increases. 
	}
	\label{fig-Lspec}
\end{figure}

These stability conclusions are corroborated by numerical computations of the spectrum for the stability problem \eqref{eq:NLS_stab}. Figure \ref{fig-Lspec} shows the spectrum for various values of $\om$ near the first critical point $\om = \om_1 \approx 0.592$. The condition $\partial_\om Q(\vp_\om) = 0$ is equivalent to a pair of eigenvalues crossing from the imaginary axis to the real axis (or vice versa) through the origin. Therefore we expect such eigenvalue zero-crossings at the critical points $\om_1, \om_2 \in (0,\gm)$ of the map $\om \mapsto Q(\vp_\om)$. These are confirmed numerically, and are indicated by the vertical lines in the top panel of Figure \ref{fig-lm2-Q}.

Let us now elaborate the behavior of eigenvalues for the spectral stability problem \eqref{eq:NLS_stab}. From the symmetry (\ref{nls-sym}), we obtain two eigenvectors in the kernel of the stability problem \eqref{eq:NLS_stab}:
\begin{align*}
\begin{pmatrix}
0 & \mL_- \\
-\mL_+ & 0
\end{pmatrix}
\begin{pmatrix}
\partial_x \varphi_{\omega} \\ 0
\end{pmatrix}
= 
\begin{pmatrix}
0 \\ 0
\end{pmatrix} \qquad 
\mbox{\rm and} \qquad 
\begin{pmatrix}
0 & \mL_- \\
-\mL_+ & 0
\end{pmatrix}
\begin{pmatrix}
0 \\ \varphi_{\omega}
\end{pmatrix}
= 
\begin{pmatrix}
0 \\ 0
\end{pmatrix}.
\end{align*}
The zero eigenvalue is generally of the quadruple algebraic multiplicity 
due to two generalized eigenvectors in the generalized kernel of the stability problem \eqref{eq:NLS_stab}. To obtain the generalized eigenvectors, 
we consider a two-parameter family of standing and traveling wave solutions 
in the form 
$$
\psi(x,t)=e^{i \omega t} \phi_{\omega,c}(x-ct),
$$
where $\omega \in (0,1)$, $c \in (-c_0,c_0)$ for a small $c_0 > 0$, and the profile $\phi_{\omega,c}$ is a solution of the complex second-order equation 
\begin{equation}
\label{ode-2-param}
(1-|\phi|^2) \phi'' + |\phi|^2 \phi = \omega \phi + i c \phi',
\end{equation}
with $\gamma = 1$. The profile $\varphi_{\omega} \equiv \psi_{\omega,c=0}$ has even parity, so that 
$\mathcal{S}_+ = (1-\varphi_{\omega}^2)^{-1} \mathcal{L}_+$ is invertible on the subspace of even functions in $L^2(\R)$ and $\mathcal{S}_- = (1-\varphi_{\omega}^2)^{-1} \mathcal{L}_-$ is invertible on the subspace of odd functions in $L^2(\R)$. As a result, the family 
$\{ \phi_{\omega,c} \}_{\omega \in (0,1), c \in (-c_0,c_0)}$ of solutions of the second-order equation (\ref{ode-2-param}) is 
smooth for some $c_0> 0$. Taking derivatives of (\ref{ode-2-param}) 
in $\omega$ and $c$, we obtain 
$$
\mathcal{L}_+ \partial_{\omega} \varphi_{\omega} = - \varphi_{\omega},\qquad 
\mathcal{L}_- \partial_{c} \phi_{\omega,c} |_{c=0} = - \partial_x \varphi_{\omega}.
$$
As a result, the generalized eigenvectors of the stability problem \eqref{eq:NLS_stab} are given by 
\begin{align*}
\begin{pmatrix}
0 & \mL_- \\
-\mL_+ & 0
\end{pmatrix}
\begin{pmatrix}
0 \\ -\partial_c \phi_{\omega,c} |_{c=0}
\end{pmatrix}
= 
\begin{pmatrix}
\partial_x \varphi_{\omega} \\ 0
\end{pmatrix} \qquad 
\mbox{\rm and} \qquad 
\begin{pmatrix}
0 & \mL_- \\
-\mL_+ & 0
\end{pmatrix}
\begin{pmatrix}
\partial_{\omega} \varphi_{\omega} \\ 0
\end{pmatrix}
= 
\begin{pmatrix}
0 \\ \varphi_{\omega}
\end{pmatrix}.
\end{align*}
In addition to the quadruple zero eigenvalue 
and the continuous spectrum on 
$$
\{ i \beta : \quad \beta \in (-\infty,-\omega] \cup [\omega,\infty) \},
$$ 
the spectral problem \eqref{eq:NLS_stab} features a bifurcating eigenvalue pair which is responsible for the switches in stability of the solitary waves.

This eigenvalue pair, shown in the top panel of Figure \ref{fig-lm2-Q} and also in Figure \ref{fig-Lspec}, emerges from the continuous spectrum at $\omega \approx 0.443$ and moves along the imaginary axis towards the origin as $\om$ increases, eventually crossing through to the real axis at $\om = \om_1 \approx 0.592$ and rendering the bright solitons unstable. The real eigenvalue pair later crosses back to the imaginary axis at $\om = \om_2 \approx 0.843$, and finally at $\omega \approx 0.912$ the eigenvalue pair disappears into the continuous spectrum. The bright solitons are thus unstable for $\om \in (\om_1, \om_2)$, and are
spectrally stable for $\om \in (0,\om_1] \cup [\om_2, \gm)$. 

Finally, we perform the following experiments to investigate the dynamical stability of bright solitons. Starting with an initial soliton $\vp_{\om_0}$ with $\omega_0 \in (0,\gm)$, we make a small perturbation and evolve the solution up to a large time $T=500$. The results shown here use generic Gaussian perturbations centered at the origin, but we have performed more experiments using different types of perturbations, including e.g. ones with sign changes, and the overall conclusions are the same as those below. We do not consider here the critical points $\om_1$ and $\om_2$ for which $\partial_\om Q(\vp_\om) = 0$. Bright solitons for such critical cases are known to be nonlinearly unstable for generic Hamiltonian systems with the $U(1)$ symmetry~\cite{CP2003,PKA96}. 

If the perturbation to the bright soliton $\vp_{\om_0}$ is sufficiently small, we expect the solution to converge, up to radiation, to a stable bright soliton. Hence the solution is expected to be of the form
\begin{align}
u(x,t) = e^{i\om_f t}\vp_{\om_f}(x) + (\text{radiation}),
\label{eq:dyn_decomp}
\end{align}
for some final frequency $\om_f$. We now seek to confirm this conjecture.

After sufficient time has passed, the radiation will have dispersed significantly and will not play any role in the central region near $x=0$. Therefore we measure the quantity $c(t)=\text{Re}(u(0,t))$, which in view of \eqref{eq:dyn_decomp} is expected to behave like $\cos(\om_f t)$, and obtain its dominant frequency via the fast Fourier transform (FFT). Once $\omega_f$ is obtained, we match the spatial profile of the evolved final state $u_f(x)$ to the bright soliton $\vp_{\omega_f}(x)$. We find good agreement between the two in all cases, confirming the decomposition \eqref{eq:dyn_decomp}.

In the case of small perturbations of the spectrally stable soliton, our computations indicate that the final frequency does not change much, i.e. $\om_f \approx \om_0$. This suggests that these bright solitons are also dynamically stable. 

\begin{figure}[htb!]
	\includegraphics[width=12cm,height=8cm]{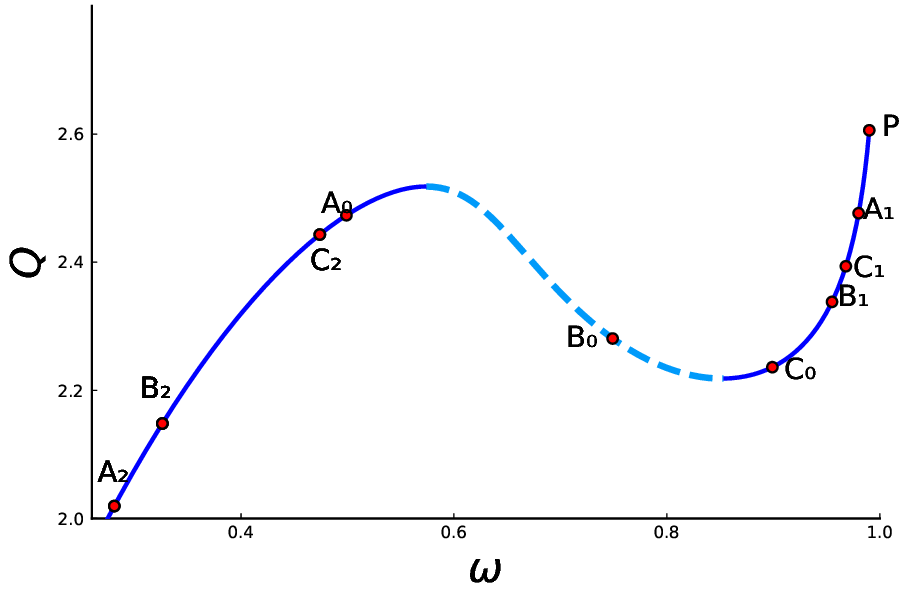}
	\caption{Initial and final frequencies shown in the $(\omega,Q)$ plane for the numerical simulations. The point $P$ represents the bright soliton with the peaked profile. The dashed line indicates the unstable branch, while the other two branches are stable.}
	\label{fig-Qvsom_evol}
\end{figure}

In the case of large perturbations, we have found that $\omega_f \neq \omega_0$ generically, nevertheless, the frequency $\om_f$ lies on one of the stable branches. In particular, it is possible to transition between the two stable branches when a large perturbation is added. For a spectrally unstable bright soliton, the perturbed solution always transforms to one of the states within the
stable branches, confirming their dynamical instability. 

\begin{figure}[htb!]
	\includegraphics[width=8cm,height=6cm]{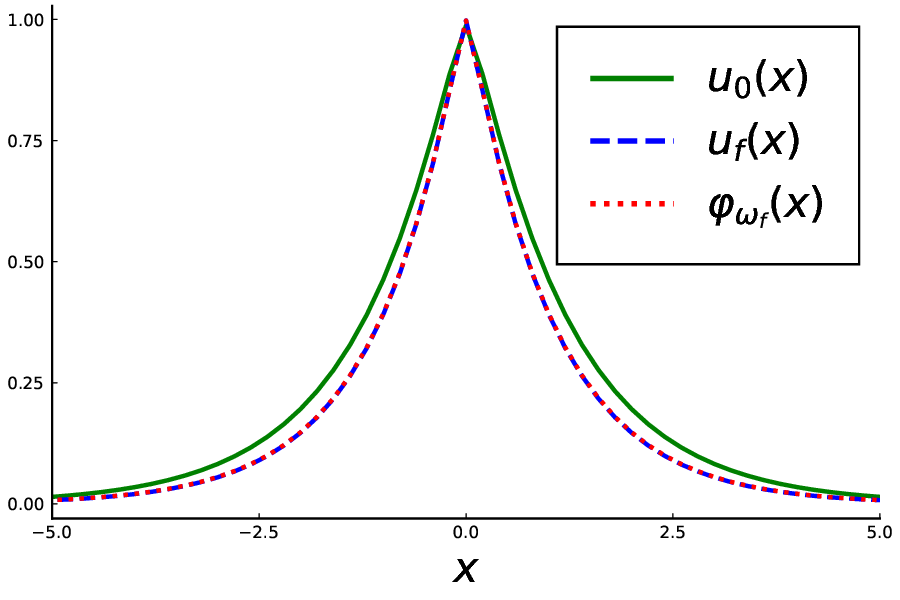}
		\includegraphics[width=8cm,height=6cm]{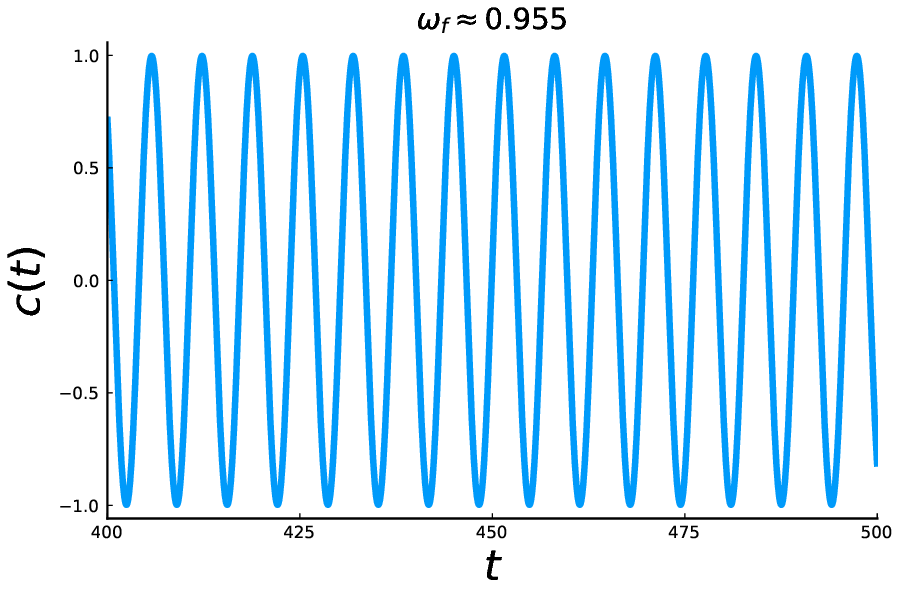}
	\caption{Left: the initial and final solution profiles of the computations 
		for the transition $B_0 \to B_1$ in Figure \ref{fig-Qvsom_evol} together with the final soliton profile. Right: oscillations near the stable profile for the final segment of numerical computations. }
	\label{fig-tolargerom}
\end{figure}

\begin{figure}[htb!]
\includegraphics[width=8cm,height=6cm]{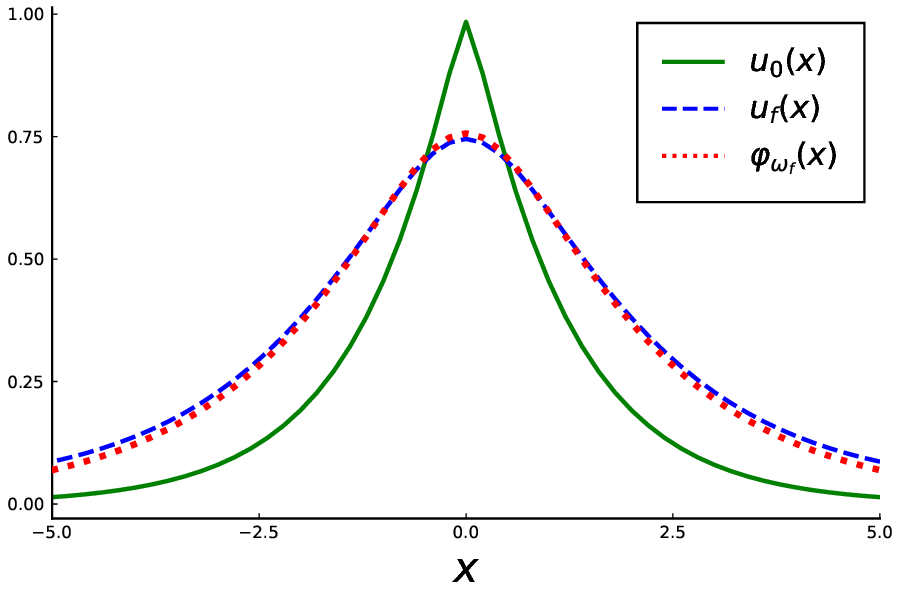}
\includegraphics[width=8cm,height=6cm]{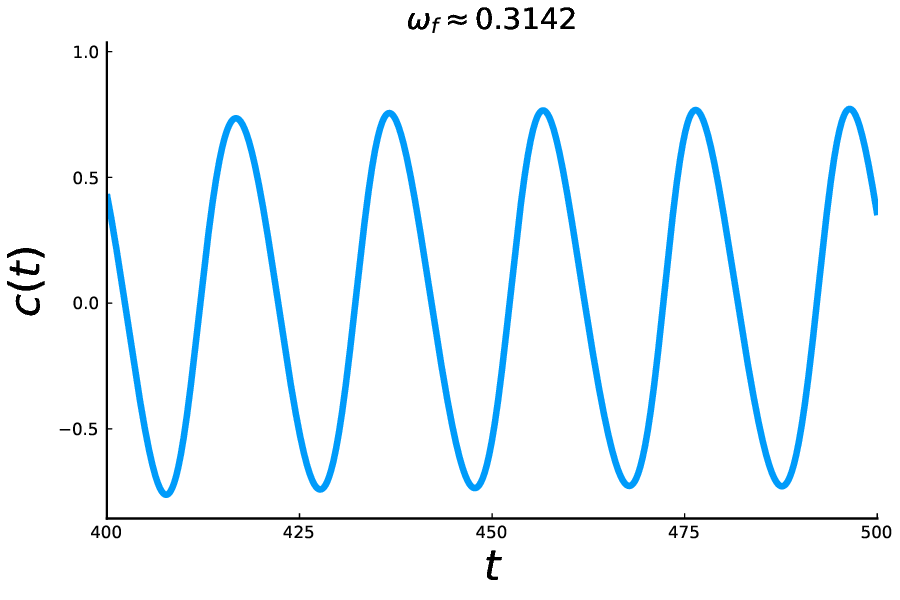}
\caption{The same as Figure \ref{fig-tolargerom} but for 
	the transition $B_0 \to B_2$ in Figure \ref{fig-Qvsom_evol}.}
\label{fig-tosmallerom}
\end{figure}

Figure \ref{fig-Qvsom_evol} summarizes in the $(\omega,Q)$ plane the outcomes of the transitions from $A_0$ and $C_0$ to either $A_1$ and $C_1$ or $A_2$ and $C_2$ in the case of large perturbations of the stable bright solitons 
and the transitions from $B_0$ to either $B_1$ or $B_2$ in the case of small perturbations of the unstable bright solitons. The two transitions are defined by Gaussian perturbations of two different signs to the initial soliton profile $\varphi_{\omega_0}$. Furthermore, details 
of the initial and final profiles are shown in Figures \ref{fig-tolargerom} and \ref{fig-tosmallerom} (left panels) 
and the final nearly periodic oscillations of the soliton amplitude 
(right panels) for the transitions $B_0 \to B_1$ and $B_0 \to B_2$.  
Note that the final spatial profiles of the bright soliton at $A_1$, $B_1$, and $C_1$ are closer to the peaked profiles which occur in the limit  $\omega \to  \gamma = 1$.

\section{Conclusion}
\label{sec-5}

In the present work we have considered a new modification of
the intensity-dependent dispersion (IDD) models, where the nonlinearly modified
dispersion competes with a local cubic nonlinearity. We argued that
this model is of interest in its own right, but also as a continuum
limit of the Salerno model (\ref{Salerno}). By using the conservation laws and by analyzing 
the stationary and spectral stability problems for this model, we showed that 
the bright solitons with smooth profiles exist if the
frequency lies within a suitable parametric interval. We have obtained 
the stability criterion for such smooth bright solitons from 
the monotonicity of the dependence of the mass on the frequency,
in line with the well-known stability criterion in the NLS-type
models. Resorting to numerical computations, we have shown that 
the smooth bright solitons are spectrally
stable for a wide range of parameters  and unstable in a narrow 
interval of frequencies. The latter were indeed
checked and identified as pertaining to spectrally unstable solutions.
Once the relevant frequency interval and its stability had been mapped,
we explored the nonlinear dynamics, confirming at first
our dynamical instability
findings but also examining the dynamical outcomes of stable solitary
waves, upon  perturbations of larger amplitudes. 

Naturally, these findings raise a number of additional questions that
are worthwhile of further investigation. In the present setting, we only
explored individual solitary waves. Yet, it would be quite interesting to
examine how the presence of IDD affects the interaction between
different solitary waves. This would not only be interesting in the
standard case of smooth solitary waves, but also in 
the limit of $\om \rightarrow \gamma$ where the profiles of bright solitons 
become peaked. 

Furthermore, the vast majority of
studies concerning IDD has been limited to one-dimensional realms.
Nevertheless, it would be particularly useful to explore the interplay
of such IDD features with, e.g., radially symmetric solutions in
higher dimensions and indeed not just ones involving standard
single-humped solitary waves, but also in more complex settings
involving vortices etc. Such studies are currently in progress and will be 
reported in future publications.

\end{document}